\documentclass[11pt]{article}

\usepackage[margin=1in]{geometry}
\usepackage{amsmath,amsfonts}
\usepackage[colorlinks=true,allcolors=black,urlcolor=blue]{hyperref}
\usepackage[font=small]{caption}
\usepackage[size=small]{subcaption}
\usepackage{isotope}
\usepackage{xcolor}

\newif\ifexporttikz
\exporttikzfalse 
\ifexporttikz
  \usepackage{tikz}
  \usepackage{pgfplots}
  \pgfplotsset{
    compat=newest,
    table/header=false,
    title style={font=\small},
    tick label style={font=\small},
    label style={font=\small},
    legend style={font=\small},
    legend cell align=left
  }
  \usetikzlibrary{spy}
  \usetikzlibrary{backgrounds}
  \usepgfplotslibrary{external}
\else
  \usepackage{tikzexternal}
  \tikzsetexternalprefix{figs/}
\fi
\newcommand{\figname}[1]{\tikzsetnextfilename{#1}}

\newcommand{\datfile}[1]{figs/#1.dat}

\definecolor{myblack}{rgb}{0.0,0.0,0.0}
\definecolor{myblue}{rgb}{0,0.4470,0.7410}
\definecolor{myred}{rgb}{0.8500,0.3250,0.0980}
\definecolor{myorange}{rgb}{0.9290,0.6940,0.1250}
\definecolor{mypurple}{rgb}{0.4940,0.1840,0.5560}
\definecolor{mygreen}{rgb}{0.4660,0.6740,0.1880}
\definecolor{mylightblue}{rgb}{0.3010,0.7450,0.9330}
\definecolor{mydarkred}{rgb}{0.6350,0.0780,0.1840}

\newcommand{\original}{\texttt{original}}
\newcommand{\hierarch}{\texttt{hierarch}}
\newcommand{\greedyres}{\texttt{greedy-res}}
\newcommand{\greedypert}{\texttt{greedy-pert}}
\newcommand{\greedyfull}{\texttt{greedy-pert-full}}
\newcommand{\newtoncorr}{\texttt{newton-corr}}

\ifexporttikz
  \pgfplotsset{OriginalStyle/.style={black,very thick,mark=*}}
  \pgfplotsset{HierarchStyle/.style={mygreen,very thick,mark=+,mark size=3pt}}
  \pgfplotsset{GreedyResStyle/.style={myred,very thick,mark=x,mark size=3pt}}
  \pgfplotsset{GreedyPertStyle/.style={myblue,very thick,mark=o,mark size=3pt}}
  \pgfplotsset{GreedyFullStyle/.style={myorange,very thick,mark=diamond,mark size=3pt}}
  \pgfplotsset{NewtonCorrStyle/.style={mypurple,very thick,mark=triangle,mark size=3pt}}
\fi

\newcommand{\xeig}[1]{x^{(#1)}}
\newcommand{\res}[1]{r^{(#1)}}
\newcommand{\lamd}[1]{\lambda^{(#1)}}
\newcommand{\Nmax}{N_{\max}}

\begin{document}

\title{\bf A greedy algorithm for computing eigenvalues\\ of a symmetric matrix}
\author{%
Taylor M.~Hernandez\textsuperscript{1},
Roel Van Beeumen\textsuperscript{2},
Mark A.~Caprio\textsuperscript{1},
Chao Yang\textsuperscript{2,}%
\footnote{Lawrence Berkeley National Laboratory,
1 Cyclotron Road, Berkeley, CA 94720. Email: \texttt{CYang@lbl.gov}.}}
\date{\small%
\textsuperscript{1}Department of Physics, University of Notre Dame,
IN, United States\\
\textsuperscript{2}Computational Research Division,
Lawrence Berkeley National Laboratory, CA, United States
}

\maketitle

\abstract{We present a greedy algorithm for computing selected
eigenpairs of a large sparse matrix $H$ that can exploit localization features
of the eigenvector.  When the eigenvector to be computed is localized,
meaning only a small number of its components have large magnitudes, the
proposed algorithm identifies the location of these components in a greedy
manner, and obtains approximations to the desired eigenpairs of $H$
by computing eigenpairs of a submatrix extracted from the corresponding
rows and columns of $H$. Even when the
eigenvector is not completely localized, the approximate eigenvectors
obtained by the greedy algorithm can be used as good starting guesses
to accelerate the convergence of an iterative eigensolver applied to
$H$.  We discuss a few possibilities for selecting important rows and columns
of $H$ and techniques for constructing good initial guesses for
an iterative eigensolver using the approximate eigenvectors returned from
the greedy algorithm. We demonstrate the effectiveness of this approach
with examples from nuclear quantum many-body calculations, many-body
localization studies of quantum spin chains and road network analysis.}


\section{Introduction}

Large-scale eigenvalue problems arise from many scientific applications.
In some of these applications, we are interested in a few eigenvalues of a large sparse symmetric matrix.
These eigenvalues may be the smallest algebraically or largest in magnitude, etc.
The dimension of these problems can be extremely large. For example, in quantum
many-body calculations, the dimension of the matrix depends on the number of particles and approximation
model parameters. It can grow rapidly with respect to the size of the problem and accuracy requirement.
For internet, social, road or traffic network analysis, the dimension of the adjacency matrix of
the network, which describes the interconnection of different nodes of the network,
depends on the number of nodes
of the network, which can increase rapidly on a daily basis.  However, these matrices are often
sparse.  Because only a small fraction of the matrix elements are nonzero, and since only a small number
of eigenpairs are desired, iterative methods are often used to solve this type of problem.
The dominant cost of these methods is in performing a sparse matrix vector
multiplication at each step of the iterative solver. For large problems, performing
this calculation efficiently on a high performance computer is a challenging task.
Not only do we need to choose an appropriate data structure to represent the
sparse matrix, we also need to develop efficient schemes to distribute the matrix
and vectors on multiple nodes or processors to overcome the single node
memory limitation and enable the computation to be performed in
parallel.

It is well known that, for some problems, the eigenvector to be computed has
localization properties, i.e., many elements of the desired eigenvector are
negligibly small~\cite{anderson,mbl}.  For example,
in quantum many-body problems, localization means that the
many-body operator of interest can be represented by a few many-body basis
functions in a small configuration space. This feature of the problem implies
that the rows and columns associated with the large elements of the
eigenvectors are more ``important" than others.
We can then effectively work with a much smaller
matrix by excluding rows and columns associated with small elements in the
eigenvector. In network analysis problems, the localization of the principal
eigenvector, i.e., the eigenvector associated with the largest (in magnitude) eigenvalue
often reveals a densely connected subgraph or cluster in the network~\cite{networkPEV}.

However, in general, we do not know which elements
of the eigenvector are small (in magnitude) in advance. In some cases, there are
efficient numerical procedures that can be used to identify
these elements, e.g., the latest work by Arnold et al.~\cite{agmonweyl,specpred}.
But these techniques may require solving another large problem such as
a large linear system of equations.
There are sometimes physical intuitions we may use
to infer which rows/columns are more important than others. For example,
in a configuration interaction approach for quantum many-body problems, the
matrix to be partially diagonalized is the representation of the Hamiltonian in a many-body
basis that consists of antisymmetric products (Slater determinants) of a set of
single-particle basis functions, e.g., eigenfunctions of a quantum harmonic
oscillator. Many-body basis functions defined by single-particle functions
associated with lower single-particle energies tend to be more ``important" than
others, although this is not always true. In a network analysis problem,
nodes with large degrees or other attributes such as centrality tend to form the center of
a densely connected subgraph, and thus are deemed more ``important" than others.

In this paper, we describe a greedy algorithm to incrementally probe
large components of a localized eigenvector to be computed.
The matrix rows and columns corresponding to these components are
extracted to construct a much smaller matrix.
The eigenvector of this small matrix is then used to obtain an
approximate eigenvector of the original matrix to be
partially diagonalized. If the approximate eigenpair is
not sufficiently accurate (the metric for measuring accuracy will be described below),
we select some additional rows and columns of the original matrix and
solve a slightly larger problem using the solution of the previous problem
as the starting guess. This procedure can be repeated recursively until the
computed eigenpair is sufficiently accurate.

For problems that are not strictly localized, i.e., many eigenvector components
are small but not zero, this approach does not completely eliminate
the need to use an iterative method to compute the desired eigenpair
of the original matrix. However, the number of iterations required to
reach convergence can be significantly reduced if a good starting guess
can be constructed from the greedy scheme.  If the submatrices
selected by the greedy algorithm are relatively small, the cost of
computing the desire eigenpairs of these smaller matrices is relatively low.
Consequently, the overall cost of the computation can be reduced.

We should note that the greedy algorithm proposed in this paper is
different from the hierarchical algorithm presented by Shao et al.~\cite{Shao18}. Instead
of using a predefined set of hierarchical configuration spaces to
construct a sequence of submatrices from which approximate eigenpairs
are computed, the greedy algorithm constructs these submatrices
dynamically using the previous approximate eigenvector to guide such
a construction.

The greedy strategy used to construct a sequence of submatrices
from which approximate eigenpairs are computed is similar to
the so-called selected configuration interaction approach
used in quantum chemistry~\cite{Tubman2016,cordsisc,cordjctc,schriber17,holmes16,cleland10,sharma17,sherrill,Harrison1991}
and the importance truncation scheme used in nuclear physics~\cite{Roth2009}.
But we would like to emphasize that the
techniques discussed here are more general. They are not restricted to
problems arising from quantum chemistry or physics. Moreover, we
describe greedy strategies in terms of matrices and vectors
instead of many-body configurations and Hamiltonians. As a result,
these strategies can potentially be applied to other applications such as sparse
principal component analysis~\cite{sparsepca}.

This paper is organized as follows. In section~\ref{sec:outline}, we
discuss the implication of eigenvector localization on the development
of an efficient iterative method for computing such an eigenvector,
and outline the general strategy for developing such an algorithm.
In section~\ref{sec:greedy}, we discuss several greedy strategies
for selecting rows and columns of the original matrix to construct
a submatrix from which approximate eigenpairs are computed and used
as a starting guess for computing the eigenpairs of the original problem.
Techniques for improving the starting guess are discussed in
section~\ref{sec:updvec}.  In section~\ref{sec:examples}, we present
some numerical examples to demonstrate the efficiency of the greedy
algorithm. Additional improvement of the algorithm is discussed in
section~\ref{sec:conclude}.

\section{Eigenvector localization and a hierarchical method for computing
localized eigenvectors}
\label{sec:outline}
Let $H \in \mathbb{R}^{n\times n}$ be the symmetric matrix to be partially diagonalized.
To simplify our discussion, let us focus on computing the algebraically
smallest eigenvalue $\lambda$ of $H$ and its corresponding eigenvector $x$.
If the desired eigenvector $x$ is localized, i.e., only a subset of
its elements are nonzero, we can reorder the elements of the
eigenvector to have all nonzero elements appear in the leading $n_1$
rows, i.e.,
\[
Px =
\begin{bmatrix}
x_1 \\
0
\end{bmatrix},
\]
where $x_1 \in \mathbb{R}^{n_1}$ and $P$ the permutation matrix associated with such a reordering.
Consequently, we can reorder the rows and columns of the matrix $H$ so that
\begin{equation}
(PHP^T) (Px) =
\begin{bmatrix}
H_1 & B \\
B^T & C
\end{bmatrix}
\begin{bmatrix}
x_1 \\
0
\end{bmatrix}
=
\lambda
\begin{bmatrix}
x_1 \\
0
\end{bmatrix}
\label{eq:PHP}
\end{equation}
holds.
To obtain $x_1$, we only need to solve the eigenvalue problem
\begin{equation}
H_1 x_1 = \lambda x_1.
\label{eq:h1eig}
\end{equation}
Even when $x$ is not strictly localized, i.e., the magnitude
of the elements in $x_1$ are significantly larger than the other elements
that are small but not necessarily zero, the solution of
\eqref{eq:h1eig} can be used to construct a good initial guess
of $x$ that can be used to accelerate the convergence of
an iterative method applied to compute the desired eigenpair of
$H$.

However, since we do not know how large the elements of $x$ are in
advance, we do not have the permutation $P$ that allows us to pick
rows and columns of $H$ to form $H_1$.

The algorithm presented in this paper seeks to identify the permutation
$P$ that allows us to construct $H_1$ incrementally so that
successively more accurate approximations to the desired eigenpair
can be obtained efficiently.
The basic algorithm we use to achieve this goal can be described as follows.
\begin{enumerate}
\item We select a subset of the indices $1,2,...,n$ denoted by $\mathcal{S}$
  that corresponds to ``important'' rows and columns of $H$.
    These rows and columns define a submatrix $H_1$ of $H$.  We will comment
    further on possible approaches to selecting $\mathcal{S}$, and therefore
    $H_1$, in Section~\ref{sec:initH1}.  For instance, in the configuration
    interaction method for solving quantum many-body eigenvalue problems, this
    subset may correspond to a set of many-body basis functions produced from
    some physically-motivated basis truncation scheme.
\item Assuming the size of $\mathcal{S}$ is
small relative to $n$, we can easily compute the desired eigenpair
$(\lambda_1, x_1)$ of $H_1$, i.e., $H_1 x_1 = \lambda_1 x_1$.
\item We take $\lambda_1$ to be the approximation to the smallest eigenvalue of $H$.
The approximation to the eigenvector of $H$ is constructed as
$\hat{x} = P^T \begin{bmatrix} x_1^T & 0 \end{bmatrix}^T$. To assess the accuracy of the computed eigenpair $(\lambda_1,\hat{x})$,
we compute the full residual $r = H \hat{x} - \lambda_1 \hat{x}$.
\item If the norm of $r$ is sufficiently small, we terminate the computation
and return $(\lambda_1,\hat{x})$ as the approximate solution. Otherwise,
we select some additional rows and columns of $H$ to augment $H_1$ and repeat
steps 2--4 again.
\end{enumerate}
If the eigenvector to be computed is localized, this procedure should terminate
before the dimension of $H_1$ becomes very large, assuming that we can identify the most important rows and columns of $H$ in some way.

We should note that we permute $H$ so that $H_1$ becomes
the leading principal submatrix merely for convenience. Such a permutation
makes our analysis more clear. But in a practical computational procedure,
as long as we can identify and access rows and columns of $H$ that
form the submatrix $H_1$, we do not need to explicitly move them to the
leading rows and columns of $H$.

\section{Greedy algorithms for detecting localization}
\label{sec:greedy}

Without loss of generality, we take $\mathcal{S}$ to be the leading $n_1$
rows and columns of $H$ so that we can partition $H$ as
\begin{equation}
H =
\begin{bmatrix}
H_1 & B \\
B^T & C
\end{bmatrix}.
\label{eq:Hpart}
\end{equation}
We now discuss how to select additional ``important'' rows and columns outside of the subset $\mathcal{S}$ to obtain a more accurate approximation of the
desired eigenvector of $H$.

\subsection{Residual based approach}
\label{sec:res}

Suppose $(\lambda_1, x_1)$ is the computed eigenpair of the submatrix $H_1$
that serve as an approximation to the desired eigenpair $(\lambda,x)$.
By padding $x_1$ with zeros to form
\begin{equation}
\hat{x} =
\begin{bmatrix}
x_1 \\
0
\end{bmatrix},
\label{eq:xhat}
\end{equation}
we can assess the accuracy of the approximate eigenvector $\hat{x}$
in the full space by computing its residual
\begin{equation}
r = H \hat{x} - \lambda_1 \hat{x}
=
\begin{bmatrix}
0 \\
B^T x_1
\end{bmatrix}
\equiv
\begin{bmatrix}
0 \\
r'
\end{bmatrix}.
\label{eq:residual}
\end{equation}

A first greedy scheme for improving the accuracy of $x_1$ is to select
$k$ row and column indices in $\{1,2,...,n\} \setminus \mathcal{S}$ that
correspond to components of $r'=B^Tx_1$ with the largest magnitude.
These indices, along with $\mathcal{S}$, yield an augmented $H_1$
from which a more accurate approximation to $(\lambda, x)$ can be obtained.

\subsection{Perturbation analysis based approach}
\label{sec:pert}

It is possible that a component of $r'$ is large in magnitude even though the
magnitude of the corresponding component in the eigenvector $x$ is relatively
small, or \textit{vice versa}. Therefore, instead of selecting row and column
indices that correspond to the components of the largest magnitude within $r'$,
it may be that a better selection can be made by estimating the magnitude of the
components of $x$ whose corresponding indices are outside of $\mathcal{S}$, and
then selecting the row and column indices that correspond to these estimated
largest elements.

To do that, let us modify the $j$th component of the zero block of
$\hat{x}$ in \eqref{eq:xhat} and assume the vector
\begin{equation}
\tilde{x} =
\begin{bmatrix}
x_1 \\
\gamma e_j
\end{bmatrix}
\label{eq:xpert1}
\end{equation}
is a better approximation to the eigenvector $x$ than $\hat{x}$ defined
in \eqref{eq:xhat}, with the corresponding eigenvalue approximation
$\tilde{\lambda} = \lambda_1 + \delta$, where $\delta$ is the correction to the eigenvalue,
and $e_j$ is the $j$th column of the $(n-n_1)\times (n-n_1)$ identity matrix.

Substituting \eqref{eq:xpert1} and $\tilde{\lambda} = \lambda_1 + \delta$ into
$Hx = \lambda x$ and examining the
$(n_1+j)$th row of the equation yields
\begin{equation}
e_j^T B^T x_1 + \gamma e_j^T C e_j  = (\lambda_1 + \delta ) \gamma.
\label{eq:pertz}
\end{equation}

If $(\lambda_1,\hat{x})$ is a good approximation to $(\lambda,x)$, i.e., both $|\delta|$ and $|\gamma|$ are relatively small,
we can drop the second order correction term $\delta \gamma$ and rearrange the equation to obtain
\begin{equation}
 (\lambda_1 - e_j^T C e_j) \gamma \approx e_j^T B^T x_1.
\end{equation}
As a result, the $(n_1+j)$th component of $x$ can be estimated to be
\begin{equation}
\gamma \approx \frac{e_j^T B^T x_1}{\lambda_1 - C_{j,j} },
\label{eq:gamma}
\end{equation}
where $C_{j,j} = e_j^T C e_j$ is the $j$ diagonal element of the matrix
$C$.

The magnitude of this quantity $\gamma$ in \eqref{eq:gamma}, the perturbation
analysis estimate for the $(n_1+j)$th eigenvector component, is then taken to
provide an estimate for the importance of the corresponding row and column of
$H$ in a greedy selection approach. We should note that
this approach is not new and has appeared in, for example, ~\cite{Harrison1991}
and ~\cite{holmes16}.  If we compare with the corresponding
component $e_j^Tr'=e_j^T B^Tx_1$ of the residual vector $r$, calculated above in
\eqref{eq:residual} to provide an estimate of the importance of this row and
column of $H$ in the residual based greedy selection approach, we see that the
quantities used to estimate the importance of a row and column in the two
approaches only differ by a scaling factor $|\lambda_1 - C_{j,j}|^{-1}$.

In \eqref{eq:xpert1}, we limit the perturbation to exactly one component in
the zero block of $\hat{x}$ in \eqref{eq:xhat}. This is the approach taken
in references~\cite{Harrison1991,Tubman2016}. We will refer to this type of
perturbation as \emph{componentwise perturbation}.

It is conceivable that perturbing several
components in this block may result in a better approximation of $x$.
In the extreme case, all components of the zero block can be perturbed to
yield a better approximation. In that case, we can express
the perturbed approximation to the desired eigenvector as
\begin{equation}
\tilde{x} =
\begin{bmatrix}
x_1 \\
z
\end{bmatrix}.
\label{eq:xpert2}
\end{equation}

Substituting \eqref{eq:xpert2} into $Hx = \lambda x$ and examine
the second block of the equation yields
\begin{equation}
B^T x_1 + C z = (\lambda_1 + \delta) z.
\label{eq:correq}
\end{equation}
Again, if we drop the second order correction term $\delta z$ and
rearrange the equation, we obtain
\begin{equation}
z \approx (\lambda_1 I - C)^{-1} B^T x_1.
\label{eq:fullz}
\end{equation}
From \eqref{eq:fullz} we can see that a full correction of the zero
component of $\hat{x}$ requires solving a linear equation with
the shifted matrix $\lambda_1 I - C$ as the coefficient.
This is likely to be prohibitively expensive because the dimension
of $C$ is assumed to be much larger than the dimension of $H_1$.
However, because all we need is the magnitudes of the components of $z$
relative to each other, which we will use to select the next set of rows and
columns of $B$ and $C$ to be included in $H_1$, we do not necessarily need to
solve the linear equation accurately. We will refer to this type of
perturbation as \emph{full perturbation}.

There are a number of options to obtain an approximate solution to
\eqref{eq:correq}.  One possibility is to use an iterative solver
such as the minimum residual (MINRES) algorithm~\cite{minres}, and perform
a few iterations to obtain an approximation to $z$.
We should note that eigenvalues of $C$ are larger than
the smallest eigenvalue of $H$. However, the approximation
to the smallest eigenvalue, $\lambda_1$, may not satisfy this
condition, i.e., $C-\lambda_1 I$ is not guaranteed to be positive definite.
Therefore, it is better to use MINRES instead of the conjugate gradient
algorithm to solve~\eqref{eq:correq}.
Another possibility is to approximate the
matrix $C$ by another matrix that is much easier to invert.
For example, if $C$ is diagonally dominant, we can replace $C$ with
a diagonal matrix $D$ that contains the diagonal of $C$. This approach
will yield the same selection criterion as that provided by~\eqref{eq:gamma}.
When $C$ is not diagonal dominant, we may also include a few
subdiagonal and superdiagonal bands to form a banded matrix approximation
to $C$. Another possibility is to replace $C$ with a block diagonal
matrix $G$ with relatively small diagonal blocks. This approach corresponds to
perturbing a few rows of the zero block of $\hat{x}$ at a time.
In this approach, it is important to block rows and columns of $C$ in
such a way that $C = G+E$ for some matrix $E$ that is relatively small
(in a matrix norm).

\subsection{The initial choice of $H_1$}
\label{sec:initH1}

The success of the greedy algorithm outlined in section~\ref{sec:outline} and
detailed in sections~\ref{sec:greedy} and~\ref{sec:updvec} depends, to a
large degree, on the initial choice of rows and columns of $H$ we select
to form the initial $H_1$. Such a choice is generally application dependent.
For example, for molecular and nuclear configuration interaction calculations,
rows and columns of $H$ that are associated with low excitation
Slater determinants, i.e., Slater determinants that consist of
single particle basis functions labeled with low quantum numbers,
are often good choices. For network analysis problems, a good choice
of rows and columns of $H$ are those associated with nodes in the network
with the largest degrees and their neighbors. For some problems, it may be possible to
start with a random selection of the rows and columns.

The number of rows and columns to be selected is also problem dependent.
Choosing a larger number of rows and columns is likely to improve
the overall convergence of the greedy algorithm and ensure the method
to converge to the desired eigenpair. On the other hand, when too many
rows and columns are chosen, the cost of computing the desired eigenpair
of $H_1$ may be close to that of computing the desired eigenpair of
$H$, which may defeat the purpose of developing the greedy algorithm.

\section{Updating the Eigenvector Approximation}
\label{sec:updvec}

Once new row and column indices have been selected using the criteria
discussed in the previous section, we update $H_1$ by including the additional
rows and columns of $B$ and $C$ specified by the new row and column indices.
We then compute the desired eigenvalue and the corresponding eigenvector
of the updated $H_1$.

Since we already have the approximate eigenvector $x_1$ associated with
the previous $H_1$, we hope to obtain the new approximation quickly
by using an iterative method that can take advantage of a good
starting guess of the desired eigenvector.

In this paper, we consider both the Lanczos method~\cite{lanczos}, which
extracts approximate eigenpairs from the Krylov subspace
\[
\mathcal{K}(H_1,v_0) = \mathrm{span}\left\{v_0, H_1 v_0, H_1^2 v_0, \cdots, H_1^{m-1} v_0\right\},
\]
where $v_0$ is the starting guess of the desired eigenvector,
and the locally optimal block preconditioned conjugate gradient (LOBPCG)
method~\cite{lobpcg}. In the LOBPCG method, the approximate eigenvector $\xeig{j}$ is
updated successively according to the following updating formula
\[
\xeig{j+1} = \alpha \xeig{j} + \beta P \res{j} + \eta \xeig{j-1},
\]
where $\res{j} = H_1 \xeig{j} - \lamd{j} \xeig{j}$ is the
residual associated with the approximate eigenpair $(\lamd{j},\xeig{j})$,
$P$ is a properly chosen preconditioner, and the scalars $\alpha$, $\beta$, and $\eta$
are chosen to minimize
the Rayleigh quotient $\langle \xeig{j+1}, H_1 \xeig{j+1}\rangle$,
subject to the normalization constraint $\langle \xeig{j+1}, \xeig{j+1}\rangle = 1$.
In addition to its ability to accelerate convergence by incorporating
a preconditioner $P$ when one is available, the LOBPCG method can also take
advantage of approximations to several eigenvectors simultaneously.
However, in this paper, we will focus on computing the lowest eigenvalue
of $H$ and its corresponding eigenvector.

There are a number of ways to choose the starting guess for both
the Lanczos method and the LOBPCG method.
The simplest approach is to construct the starting guess by
padding $x_1$ with additional zeros.
Another possibility is to pad $x_1$ with the largest components (in magnitude)
of the approximate solution $z$ defined by \eqref{eq:fullz}, especially
if \eqref{eq:fullz} is used to select the new rows and columns of $B$ and
$C$ to be included in $H_1$.  This approach may work well if
components of $x_1$ are already very close to the corresponding
components in the exact eigenvector $x$. However, if that is not the case,
we need to correct $x_1$ as well by defining $\tilde{x}$ as
\begin{equation}
\tilde{x} =
\begin{bmatrix}
x_1 + z_1 \\
z_2
\end{bmatrix},
\qquad x_1^T z_1 = 0.
\label{eq:xpert3}
\end{equation}

Substituting $\tilde{x}$ and $\lambda = \lambda_1 - \delta$ into
$Hx = \lambda x$, enforcing the $x_1^T z_1 = 0$ constraint, and dropping the second order perturbation term yields
\begin{equation}
\begin{bmatrix}
H_1   & B               & x_1\\
B^T   & C - \lambda_1 I & 0 \\
x_1^T & 0               & 0
\end{bmatrix}
\begin{bmatrix}
z_1 \\
z_2 \\
\delta
\end{bmatrix}
=
\begin{bmatrix}
0 \\
-B^Tx_1 \\
0
\end{bmatrix}.
\label{eq:z12d}
\end{equation}
Eliminating $\delta$ from \eqref{eq:z12d} and applying the
projector
\[
\begin{bmatrix}
I - x_1 x_1^T & 0 & 0 \\
0             & I & 0 \\
0             & 0 & 1
\end{bmatrix}
\]
to both sides of the equation results in
\begin{equation}
\begin{bmatrix}
\hat{H}_1 & \hat{B} \\
\hat{B}^T       & C - \lambda_1 I
\end{bmatrix}
\begin{bmatrix}
z_1 \\
z_2
\end{bmatrix}
=
\begin{bmatrix}
0 \\
-B^T x_1
\end{bmatrix},
\label{eq:correq12}
\end{equation}
where $\hat{H}_1 = (I-x_1x_1^T)(H_1 - \lambda_1 I)(I-x_1x_1^T)$ and
$\hat{B} = (I-x_1x_1^T)B$. Note that the above derivation of the correction
equation \eqref{eq:correq12} is similar to that used in the development of the Jacobi-Davidson
algorithm~\cite{jd}.

We can solve \eqref{eq:correq12} by using an iterative solver such as
the MINRES algorithm. Instead of adding $z_1$ and $z_2$ directly
to $\begin{bmatrix} x_1^T & 0 \end{bmatrix}^T$ as shown in \eqref{eq:xpert3}, we can project
$H$ into a two-dimensional subspace spanned by
\[
Q =
\left\{
\begin{bmatrix}
x_1 \\
0
\end{bmatrix},
\begin{bmatrix}
z_1 \\
z_2
\end{bmatrix}
\right\},
\]
and solving a $2\times 2$ eigenvalue problem. If $g_1$ is the eigenvector
associated with the smallest eigenvalue of the projected matrix,
the starting guess of the desired eigenvector of $H$ can be
chosen as
\[
\tilde{x} = Qg_1.
\]
We will refer to this approach of preparing the starting guess as
the \emph{Newton correction}.

\section{Numerical examples}
\label{sec:examples}

In this section, we demonstrate the effectiveness of the greedy algorithm
for computing the lowest eigenvalue of $H$ for three different applications.
The first one arises from nuclear structure calculations. The second is
concerned with computing the localized eigenvector of a model many-body
Hamiltonian that includes local interactions and a disordered
potential term. The third is related to computing
the principal eigenvector of an adjacency matrix associated with a network graph.
Our computations are performed on a Microsoft Surface Laptop 2 with an Intel Core i7-8650 chip running at 1.9Ghz, and 16GB memory. Our code is written in MATLAB and the version of MATLAB we use is
R2018a.

Before presenting the results of the numerical experiments, we first describe 2 reference calculations, the 3 different variants of the greedy algorithm to be compared, and the Newton correction approach below:
\begin{itemize}
\item \original: reference calculation by directly solving the full problem and using a random vector as initial guess.
\item \hierarch: reference calculation by exploiting the hierarchical structure of the matrix $H$, i.e., first solving the small problem, next padding the obtained small eigenvector with zeros and using it as initial guess for the full problem.
\item \greedyres: greedy algorithm which uses the residual based approach for selecting the row and column indices to augment $H$ with.
\item \greedypert: greedy algorithm which uses the componentwise perturbation analysis based approach \eqref{eq:gamma} for selecting the row and column indices to augment $H$ with.
\item \greedyfull: greedy algorithm which uses the full perturbation analysis based approach \eqref{eq:fullz} for selecting the row and column indices to augment $H$ with.
\item \newtoncorr: Newton correction approach \eqref{eq:xpert3} for updating the eigenvector and initial guess.
\end{itemize}

\subsection{Nuclear Configuration Interaction}
\label{sec:examples:nuclear}

The matrix $H$ to be partially diagonalized in this example is the nuclear many-body
Schr\"{o}dinger Hamiltonian for the nucleus of a lithium atom, in particular, of
the isotope $\isotope[6]{Li}$, for which the nucleus consists of 3 protons and 3
neutrons. The matrix approximation to the nuclear Schr\"{o}dinger Hamiltonian
operator is constructed on the so-called configuration interaction space,
spanned by a set of many-body basis functions.

Each of these many-body basis functions is a Slater determinant of a set of
eigenfunctions of a 3D harmonic oscillator. These single-particle eigenfunctions
are indexed by a set of quantum numbers $\{n(a), l(a), j(a), m(a)\}$, for each
nucleon $a$, and is associated with number $N(a)=2n(a)+l(a)$ of oscillator
quanta~\cite{suhonen2007:nucleons-nucleus}.  In the nuclear physics
applications, the selection of Slater determinants for the configuration space
is often done by specifying a limit on the sum of the oscillator quanta
$N_\mathrm{tot}=\sum_a [2n(a)+l(a)]$ (some additional constraints are imposed on
the quantum numbers to ensure appropriate symmetry
properties)~\cite{barrett2013:ncsm}.  This limit on the oscillator quanta is
often expressed in terms of a cutoff parameter $\Nmax$ indicating the limit on
the number of quanta permitted \textit{above} the minimal number $N_0$ possible
(that is, consistent with the Pauli principle, or antisymmetry of Slater
determinants) for that nucleus: then the many-body basis function is restricted
to $N_\mathrm{tot}\leq N_0+\Nmax$.  This constraint defines a truncation
relative to the full configuration space, defined by all possible Slater
determinants that can be generated, from a given set of harmonic oscillator
eigenfunctions.  The larger the $\Nmax$, the larger the dimension of the
matrix approximation $H$ to the Hamiltonian, and the higher the cost to obtain the desired
eigenpairs of $H$.

\begin{figure}[htbp]
 \centering
 \captionsetup{justification=raggedright}
 \hfill
 \begin{subfigure}[t]{.5\textwidth}
    \centering
    \includegraphics[width=0.9\textwidth]{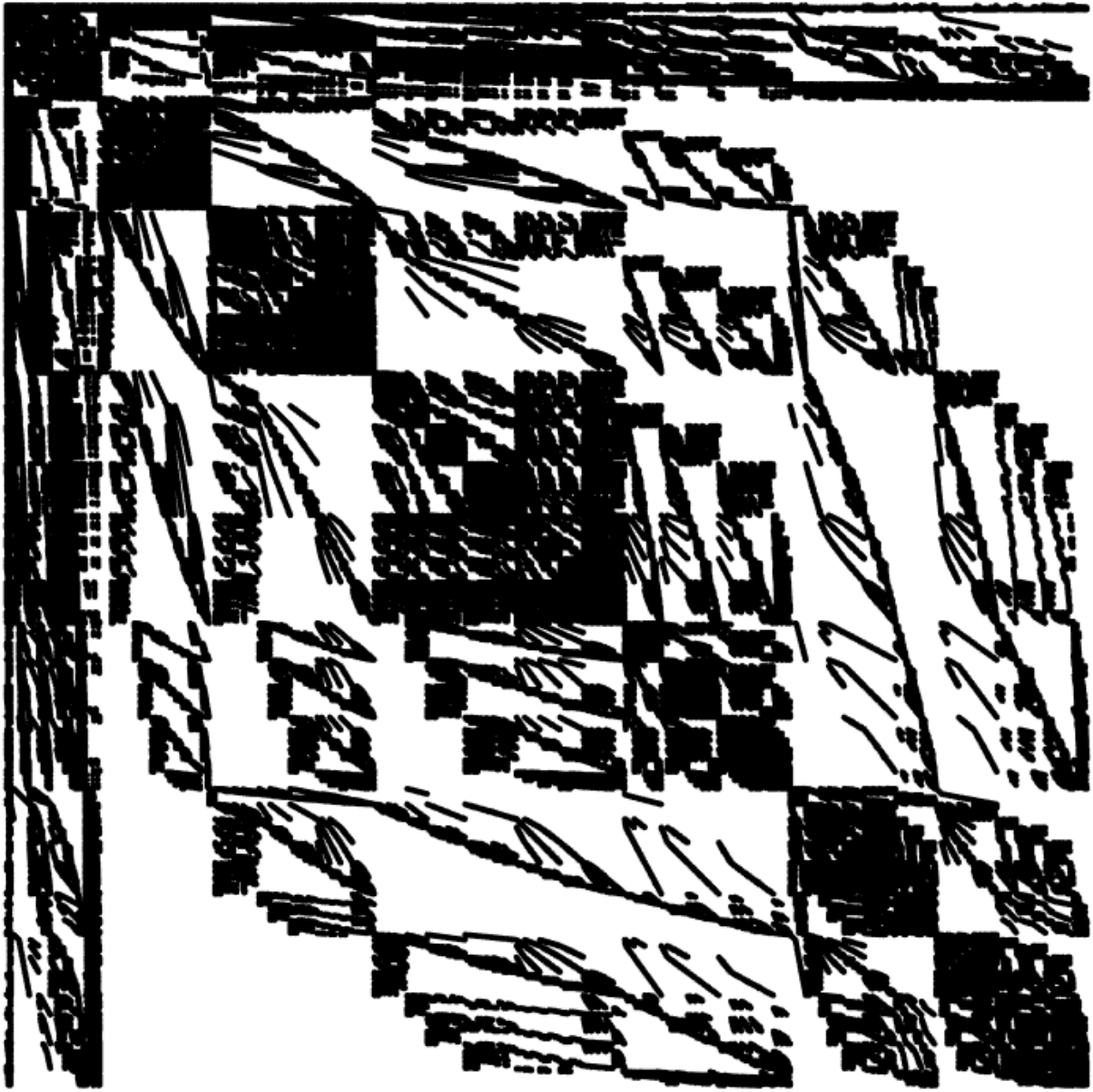}
    \caption{The sparsity pattern of the $\isotope[6]{Li}$ Hamiltonian.}
 \end{subfigure}%
 \hfill
 \begin{subfigure}[t]{.5\textwidth}
    \centering
    \figname{nci-eigvec}%
    \begin{tikzpicture}
    \begin{axis}[
      width=0.95\textwidth,
      height=0.95\textwidth,
      xmin=0, xmax=197900,
      scaled x ticks=false,
      ymin=0, ymax=0.5,
      xtick={0,50000,100000,150000,197882},
      xticklabels={0,{50,000},{100,000},{150,000},{197,882}},
      xlabel={$i$},
      ylabel={$|x(i)|$},
    ]
    \addplot+[ycomb,mark=none,myblack,line width=1pt] table {\datfile{nci-eigvec}};
    \end{axis}
    \end{tikzpicture}
    \caption{Magnitude of each element of first eigenvector.}
 \end{subfigure}%
\caption{The sparsity structures of the $\isotope[6]{Li}$ Hamiltonian in the $\Nmax = 6$ configuration space and its first
eigenvector.}
\label{fig:Li6mat}
\end{figure}
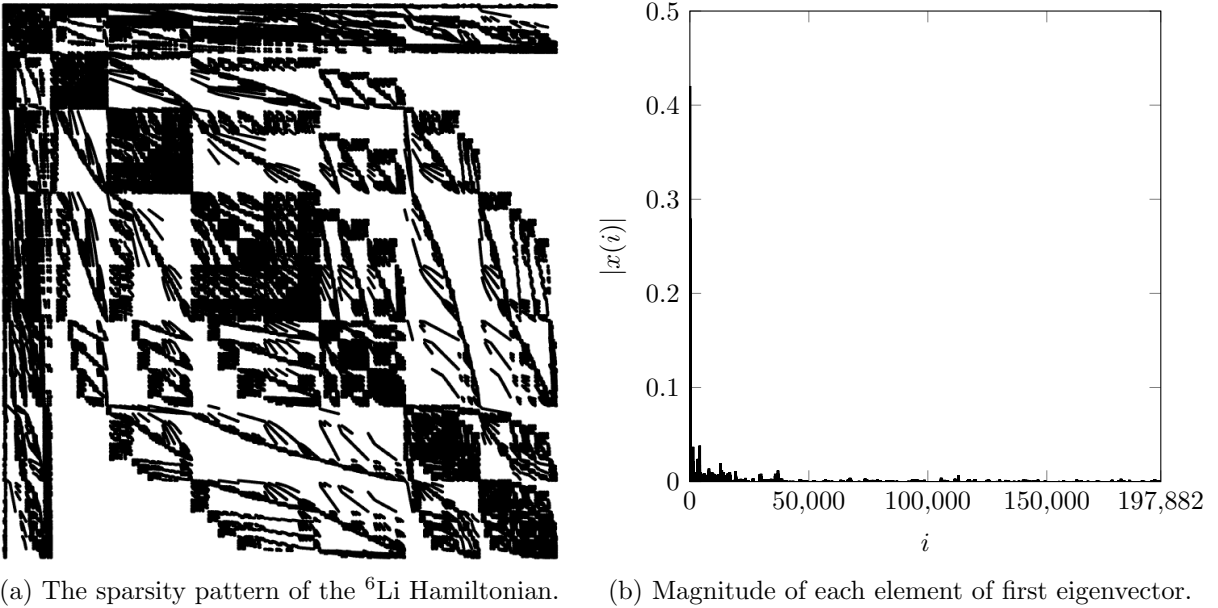

We consider only two-body potential interactions. The finite
dimensional Hamiltonian constructed from a truncated configuration
space is sparse.  Figure~\ref{fig:Li6mat}(a) shows the nonzero matrix
element pattern of $H$ for the $\Nmax=6$ truncation level.
The dimension of this matrix is 197,882. A lexicographical ordering
of the many-body (Slater determinant) basis by its single particle quantum
numbers presented in~\cite{Sternberg2008} is used. This ordering preserves
the $\Nmax$ model truncation hierarchy. Under such a ordering scheme, the leading $800 \times 800$
principal submatrix of $H$ corresponds to the Hamiltonian truncated
with $\Nmax = 2$.

As a reference, we use the LOBPCG algorithm to
compute the lowest eigenvalue and its eigenvector of $H$ for $\Nmax = 6$.
For simplicity, no preconditioner is used here. However,
an effective preconditioner such as the block diagonal preconditioner
presented in~\cite{Shao18} can be used.
We plot the magnitude of its components in Figure~\ref{fig:Li6mat}(b).
As we can see, many of these components are small.

We first compare the perturbation analysis based greedy algorithm (\greedypert) to the standard LOBPCG algorithm (\original) and the hierarchical approach (\hierarch).
It is clear from Figure~\ref{fig:Li6mat}(b) that the largest components
(in magnitude) of the eigenvector appear in the leading portion of the vector
that correspond to the configuration space defined by a small $\Nmax$
truncation level.  Therefore, we take the leading $800\times 800$ principal
submatrix ($\Nmax=2$) as the starting point of both the hierarchical and greedy algorithm and compute its smallest eigenvalue and corresponding eigenvector using the LOBPCG algorithm.

For the greedy algorithm,
we then use the $\gamma$ value defined in \eqref{eq:gamma} to
select additional rows and columns to augment the matrix $H_1$.
We first select all rows and columns with $|\gamma|$ greater than a
threshold of $\tau = 5\times 10^{-3}$. The total number
of selected rows (and columns) is 62. Next, we compute the lowest
eigenvalue and corresponding eigenvector of this
$862\times 862$ matrix $H_1$ using the zero padded eigenvector
of the previous $H_1$ as the starting guess, and perform the perturbation
analysis again to select additional rows and columns. Using
the threshold value of $\tau = 5 \times 10^{-4}$ yields an augmented
matrix $H_1$ of dimension 8004. Although it is possible to continue
this process by using a lower threshold to select additional
rows and columns to further augment $H_1$, a slightly lower threshold
actually results in a significant increase in the number of
new rows and columns to be included in $H_1$. This makes it
costly to compute the desired eigenpair of $H_1$ even when
a zero padded eigenvector of the previous $H_1$ is used as
the starting guess. We believe this is because the eigenvector associated
with the smallest eigenvalue of $H$ is not completely localized,
since more than 51\% of the components of the eigenvector
have magnitude less than $10^{-4}$ and less than 10\% of them are
less than $10^{-5}$ in magnitude.
Therefore, we stop the greedy selection of additional rows and columns
when the dimension of $H_1$ reaches 8004, and use the eigenvector
associated with the smallest eigenvalue of this problem as the starting
guess to compute the ground state of $H$, after it is padded with zeros.

\begin{figure}[tp]
\centerline{%
\figname{nci-conv-original-vs-greedy}%
\begin{tikzpicture}
\begin{semilogyaxis}[
  width=0.7\textwidth,
  height=0.5\textwidth,
  xmin=0, xmax=90,
  ymin=5e-5, ymax=2e1,
  xlabel={LOBPCG iteration},
  ylabel={relative residual norm},
  legend pos=north east,%
  legend style={draw=none,fill=none},%
]
\addplot[gray,densely dashed] coordinates{(0,1e-4) (90,1e-4)};
\addplot[OriginalStyle] table {\datfile{nci-original}};
\addplot[HierarchStyle] table {\datfile{nci-hierarch}};
\addplot[GreedyPertStyle] table {\datfile{nci-greedy-pert}};
\legend{,\original,\hierarch,\greedypert};
\end{semilogyaxis}
\end{tikzpicture}%
}
\caption{The convergence of the LOBPCG algorithm for computing the
ground state of the $\isotope[6]{Li}$ Hamiltonian at the $\Nmax=6$ truncation
level when the initial approximation to the eigenvectors is
prepared with a greedy algorithm, a hierarchical scheme and a random
vector.\label{fig:Li6conv}}
\end{figure}
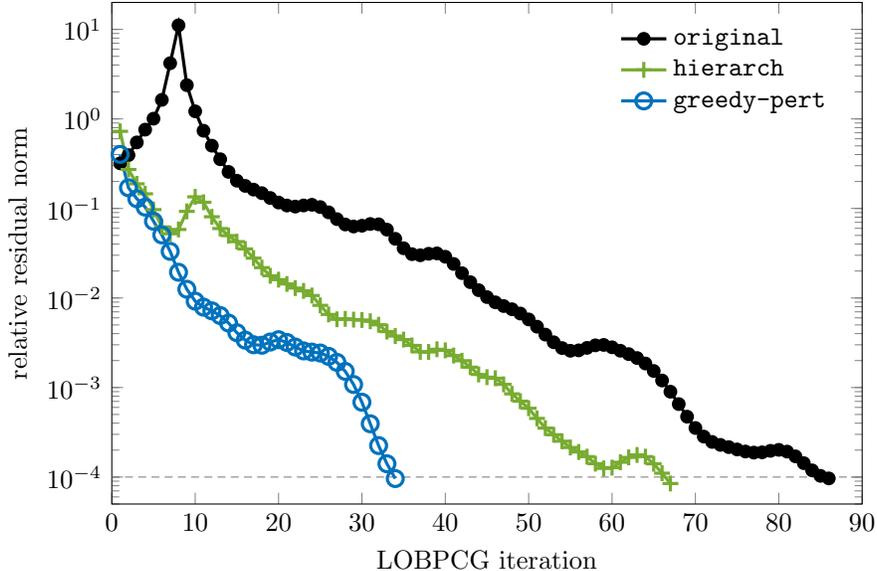

Figure~\ref{fig:Li6conv} shows the convergence history of the LOBPCG
algorithm applied to $H$ using as starting guess a random starting vector (\original), the small eigenvector of size 800 padded by zeros (\hierarch), and the zero padded eigenvector obtained by the greedy approach from the $8004\times 8004$ $H_1$ (\greedypert). We plot the relative residual norm
defined as
\begin{equation}
\| H x^{(k)} - \theta^{(k)} I \| / |\theta^{(k)}|,
\label{eq:resnrm}
\end{equation}
where $k$ is the iteration number, and $(\theta^{(k)}, x^{(k)})$ are the
approximate eigenvalues and corresponding eigenvectors obtained at the $k$th iteration.
We can see from Figure~\ref{fig:Li6conv} that the starting
vector constructed from the greedy approach enables the LOBPCG algorithm
to converge in less than half of the number of iterations required in
either the ``original'' approach and ``hierarchical'' approach.
In terms of the total wall clock time, which includes the time required
to compute eigenpairs of the sequence of $H_1$ matrices, the greedy algorithm is
2.5 times faster than the ``original'' approach, and 1.9 times faster
than the ``hierarchical'' approach.

\begin{figure}[hbtp!]
\centerline{%
\figname{nci-compare-greedy}%
\begin{tikzpicture}
\begin{semilogyaxis}[
  width=0.7\textwidth,
  height=0.5\textwidth,
  xmin=0, xmax=35,
  ymin=5e-5, ymax=1e0,
  xlabel={LOBPCG iteration},
  ylabel={relative residual norm},
  legend pos=north east,%
  legend style={draw=none,fill=none},%
]
\addplot[gray,densely dashed] coordinates{(0,1e-4) (35,1e-4)};
\addplot[GreedyResStyle] table {\datfile{nci-greedy-res}};
\addplot[GreedyPertStyle] table {\datfile{nci-greedy-pert}};
\addplot[GreedyFullStyle] table {\datfile{nci-greedy-full}};
\legend{,\greedyres,\greedypert,\greedyfull};
\end{semilogyaxis}
\end{tikzpicture}%
}
\caption{A comparison of the convergence of the LOBPCG algorithm when
it is applied to $H$ with starting vectors obtained
from greedy algorithms that use residual and perturbation analysis
respectively to select rows and columns.
All 3 methods (\greedyres, \greedypert, \greedyfull) use zero padded starting vectors.}
\label{fig:Li6compres}
\end{figure}
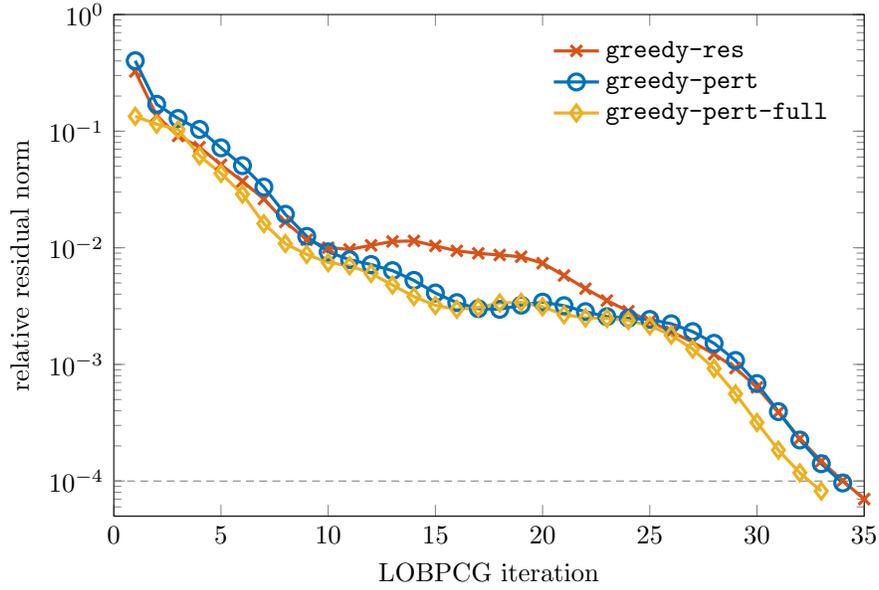

We now compare the perturbation analysis based greedy approach (\greedypert) to the residual based (\greedyres) and full perturbation analysis based (\greedyfull) greedy approaches.
Instead of using the componentwise perturbation analysis and selecting rows and
columns to be included in $H_1$ by examining the magnitude of
$\gamma$, we can examine the magnitude of the residual
$r'$ defined in \eqref{eq:residual} and choose rows and columns of
$H$ associated with elements of $r'$ that are sufficiently large in
magnitude.  By setting the threshold $\tau$ to $5\times 10^{-1}$ and
$10^{-1}$ respectively, we generate $H_1$ matrices of
similar dimensions compared to those generated from the perturbation
analysis based approach. Using the eigenvector computed from the
larger $H_1$ matrix, we are able to obtain the desired eigenpair of $H$
with nearly the same number of LOBPCG iterations as used by the
perturbation based approach as we can see in
Figure~\ref{fig:Li6compres}.

As discussed in section~\ref{sec:greedy},
the most costly linear perturbation analysis
requires (approximately) solving a linear equation of the form
given in \eqref{eq:fullz} to produce the vector $z$ that can be used
for selecting additional rows and columns.
In this example, we solve \eqref{eq:fullz} by running 10 iterations of
the MINRES algorithm.  By setting the  threshold $\tau$ to $5\times 10^{-3}$
and $10^{-3}$ respectively, we obtain $H_1$ matrices of
similar dimensions compared to those generated from the componentwise
perturbation analysis based approach. Using the eigenvector computed from the
larger $H_1$ matrix as the starting vector, we are able to obtain the desired
eigenpair of $H$ with a slightly fewer iterations as we can see in
Figure~\ref{fig:Li6compres}.  However, since we need to solve \eqref{eq:fullz},
the overall cost of this approach is actually slightly higher.

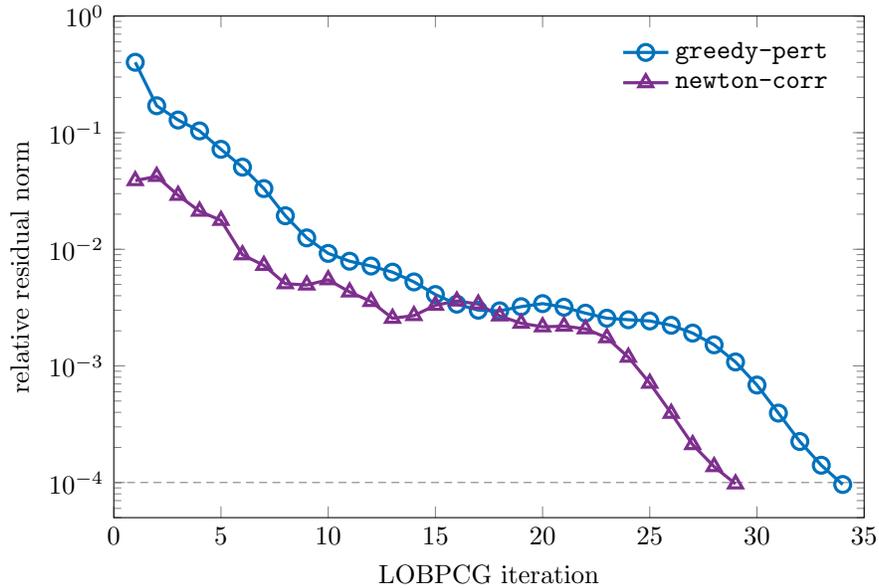
\begin{figure}[hbtp!]
\centerline{%
\figname{nci-compare-newton}%
\begin{tikzpicture}
\begin{semilogyaxis}[
  width=0.7\textwidth,
  height=0.5\textwidth,
  xmin=0, xmax=35,
  ymin=5e-5, ymax=1e0,
  xlabel={LOBPCG iteration},
  ylabel={relative residual norm},
  legend pos=north east,%
  legend style={draw=none,fill=none},%
]
\addplot[gray,densely dashed] coordinates{(0,1e-4) (35,1e-4)};
\addplot[GreedyPertStyle] table {\datfile{nci-greedy-pert}};
\addplot[NewtonCorrStyle] table {\datfile{nci-greedy-newt}};
\legend{,\greedypert,\newtoncorr};
\end{semilogyaxis}
\end{tikzpicture}%
}
\caption{A comparison of the convergence of the LOBPCG algorithm when
it is applied to $H$ with starting vectors
produced by the greedy algorithm padded with zeros (\greedypert) or
corrected by Newton's method (\newtoncorr).}
\label{fig:Li6compv0}
\end{figure}

We suggested in section~\ref{sec:updvec} that it may be more beneficial
to correct the eigenvector obtained from the small configuration
space by performing a Newton correction which requires solving
\eqref{eq:correq12}. Figure~\ref{fig:Li6compv0} shows that
such a starting vector yields a noticeable reduction in the
number of LOBPCG iterations compared to the approach that simply
constructs the initial guess by padding $x_1$ with zeros.
In this example, equation \eqref{eq:correq12} is solved by running
5 MINRES iterations. If we take into account the cost required
to solve \eqref{eq:correq12}, the overall cost of the Newton
correction approach is comparable to that used by the zero padding approach.

\subsection{Many-Body Localization}
\label{sec:examples:mbl}

In this section, we give another example that illustrates the
effectiveness of the greedy algorithm. The matrix of interest
is a many-body Hamiltonian (Heisenberg spin-1/2 Hamiltonian) associated with a disordered quantum spin chain
with $L=20$ spins and nearest neighbor interactions. The Hamiltonian
has the form
\begin{equation}
H = \sum_{i=1}^{L-1} I \otimes \cdots \otimes I \otimes H_{i,i+1} \otimes I \otimes \cdots \otimes I
    + \sum_{i=1}^L I \otimes \cdots \otimes I \otimes h_i S_i^z \otimes I \otimes \cdots \otimes I,
\label{eq:MBL}
\end{equation}
where the parameters $h_i$ are randomly generated and represent the disorder, $I$ is the 2-by-2 identity matrix, and
\[
H_{i,i+1} = S_i^x \otimes S_{i+1}^x +
            S_i^y \otimes S_{i+1}^y +
            S_i^z \otimes S_{i+1}^z
\]
is a 4-by-4 real matrix,
with $S^x$, $S^y$, and $S^z$ being spin matrices (related to the Pauli matrices by a factor of $1/2$), defined as
\[
S^x =
\frac{1}{2}
\begin{bmatrix}
0 & 1 \\
1 & 0
\end{bmatrix},
\quad
S^y =
\frac{1}{2}
\begin{bmatrix}
0 & -i \\
i & 0
\end{bmatrix},
\quad
S^z =
\frac{1}{2}
\begin{bmatrix}
1 & 0 \\
0 & -1
\end{bmatrix},
\]
respectively.
Note that the matrices $H_{i,i+1}$ are identical for all $i$,
their subscripts simply indicating the overlapping
positions in each Kronecker product.

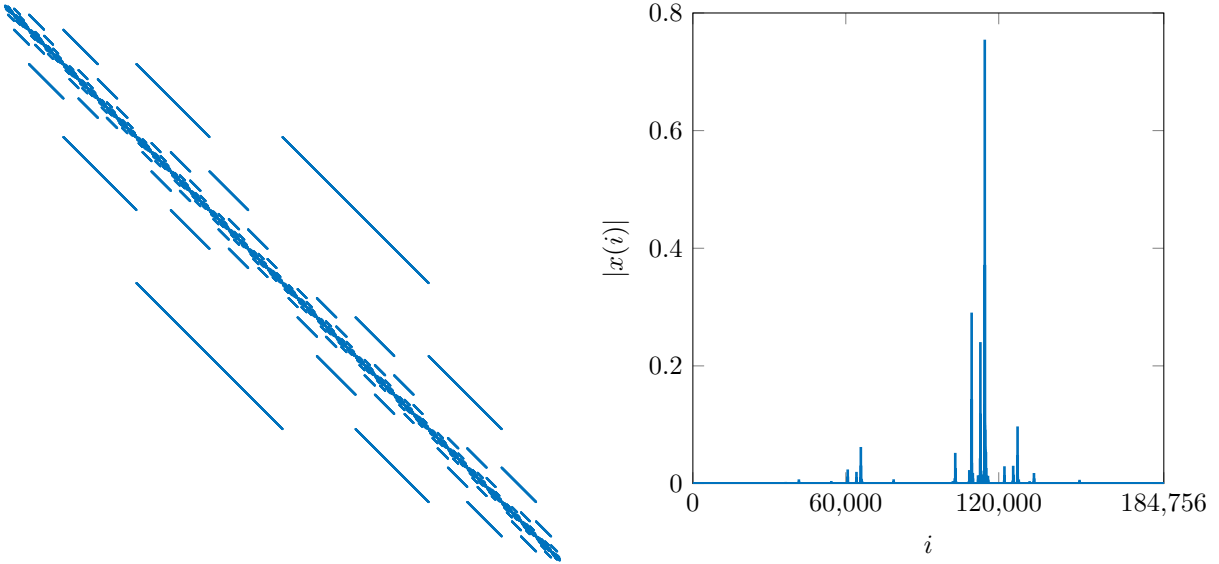
\begin{figure}[b!]
 \centering
 \hfill
   \begin{subfigure}[t]{.5\textwidth}
     \centering
     \includegraphics[width=0.9\textwidth]{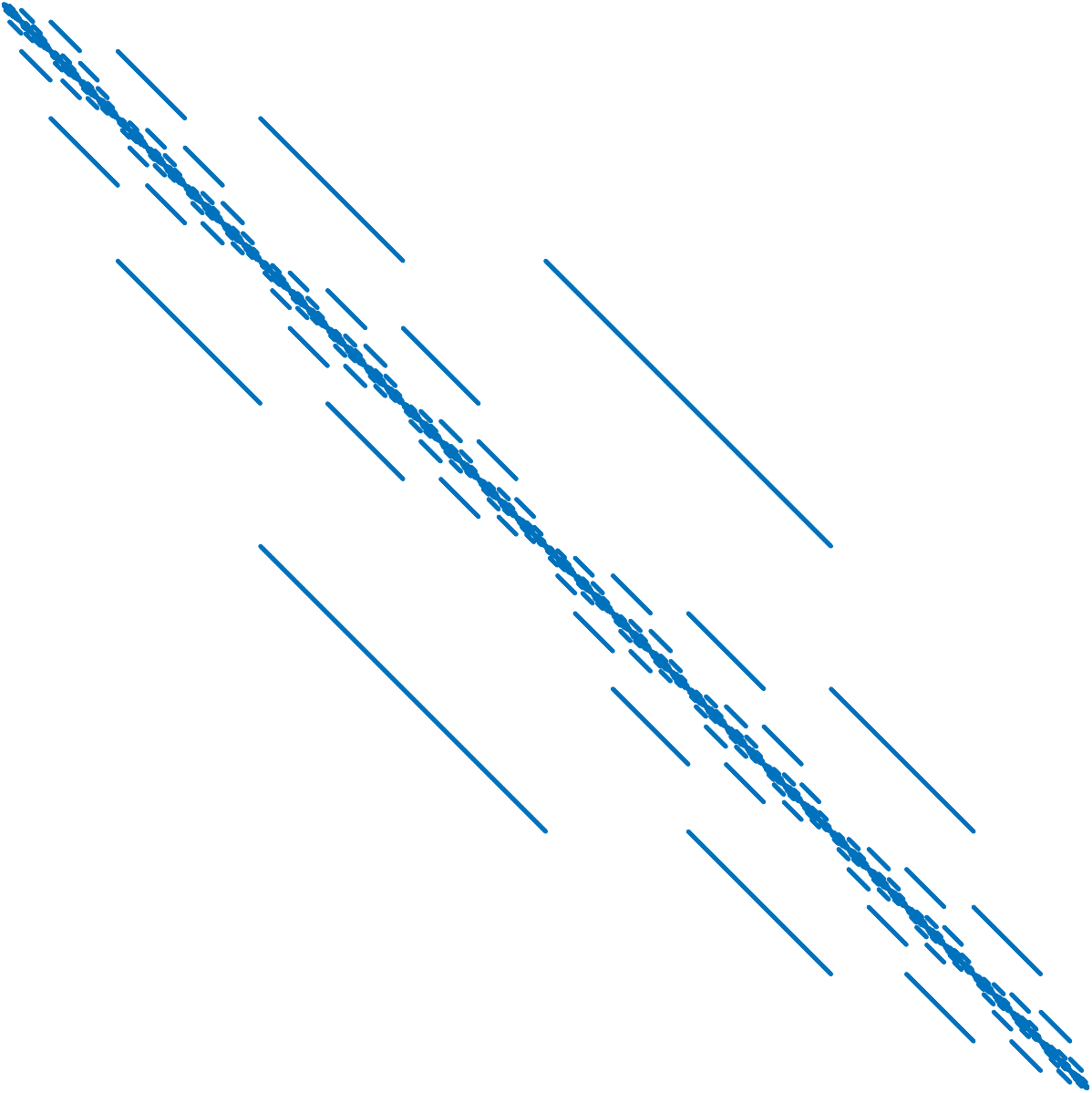}
     \caption{Sparsity pattern of the many-body Hamiltonian.}
   \end{subfigure}%
 \hfill
    \begin{subfigure}[t]{.5\textwidth}
    \centering
    \figname{mbl-eigvec}%
    \begin{tikzpicture}
    \begin{axis}[
      width=0.95\textwidth,
      height=0.95\textwidth,
      xmin=0, xmax=184756,
      scaled x ticks=false,
      ymin=0, ymax=0.8,
      xtick={0,60000,120000,184756},
      xticklabels={0,{60,000},{120,000},{184,756}},
      xlabel={$i$},
      ylabel={$|x(i)|$},
    ]
    \addplot+[ycomb,mark=none,myblue,line width=1pt] table {\datfile{mbl-eigvec}};
    \end{axis}
    \end{tikzpicture}
    \caption{Magnitude of each element of first eigenvector.}
    \end{subfigure}%
 \hfill
 \caption{The sparsity structures of the many-body Hamiltonian associated
with a Heisenberg spin chain of length 20, and its eigenvector associated
with the lowest eigenvalue.}
 \label{fig:HMBL}
\end{figure}

The matrix $H$ \eqref{eq:MBL} can be permuted into a block diagonal form.
We are interested in
the lowest eigenvalue of the largest diagonal block, which corresponds
to half the spins being up and the other half down. The sparsity
structure of this matrix, which has a dimension of $184,756$,
is shown in Figure~\ref{fig:HMBL}(a).
When the disorder $h$ is sufficiently large, the eigenvectors of $H$ exhibit a
localized feature~\cite{luitz15,hpcasia}. Figure~\ref{fig:HMBL}(b) shows the eigenvector
associated with the lowest eigenvalue.

When applying the greedy algorithm (\greedypert) to $H$, we first randomly pick 200
rows and columns of the $H$ matrix and compute the lowest eigenvalue, i.e., the ground state, and the corresponding eigenvector of $H_1$ using the
\texttt{eigs} function in
MATLAB, which implements the implicitly restarted Lanczos~\cite{arpack} or the Krylov--Schur method~\cite{stewart02}. The choice of $200$ rows and columns is arbitrary. However, when
these rows and columns are chosen randomly, the number of rows and columns has to be
sufficiently large in order to prevent the greedy algorithm from converging to
an undesired eigenpair. For this particular problem, 200 appears to be sufficiently large while still
being relatively small compared to the dimension of $H$. We repeated the experiments multiple times
with different choices of the first random 200 rows and column. All runs converge to the
ground state.

We use the first order componentwise perturbation method to seek additional
rows and columns of $H$ to add to the submatrix $H_1$ to be partially diagonalized.
To be specific, we choose rows (and columns) whose corresponding $\gamma$
value defined in \eqref{eq:gamma} is less than a threshold of $\tau=10^{-3}$
and add these to the submatrix $H_1$. We
compute the lowest eigenvalue and its corresponding eigenvector of
the augmented matrix again.

Then, rather than simply using the result to provide a starting guess vector for
a diagonalization of the full matrix $H$, we choose to repeatedly iterate the
perturbation based greedy selection process.  If no additional rows (and
columns) can be selected with the current threshold, we lower the threshold by a
factor 10.  We terminate the computation when the relative residual norm is
below $10^{-7}$.  Because this accuracy is already sufficient, we do not need to
follow up with a calculation on the full Hamiltonian, that is, using a zero
padded starting guess, as we have previously described doing at the end of the
greedy selection procedure.

Table~\ref{tab:MBLconv} shows the relative residual norm of
the approximate eigenpair, $\|r\|/|\theta|$, where $r$ is
is defined by \eqref{eq:residual}, for each of the thresholds $\tau$
used in the greedy procedure.  For each $\tau$ value, we also show
the dimension of the augmented $H_1$ right before the threshold is lowered,
and the wall clock time used to compute the desired eigenpairs
for successively augmented $H_1$ matrices generated for that particular
$\tau$ threshold.

\begin{table}[hbtp]
\centering
\begin{tabular}{|c|c|c|c|}
\hline
threshold ($\tau$) & $\|r\|/|\theta|$ & dim($H_1$) & wall clock time (sec) \\ \hline \hline
$10^{-3}$ &  $4.7\times 10^{-3}$& $986$  & $2.7\times 10^{-3}$ \\ \hline
$10^{-4}$ &  $1.0\times 10^{-3}$& $2,546$ & $2.9\times 10^{-3}$ \\ \hline
$10^{-5}$ &  $1.7\times 10^{-4}$& $4,316$ & $6.0\times 10^{-3}$ \\ \hline
$10^{-6}$ &  $2.3\times 10^{-5}$& $7,558$ & $8.5\times 10^{-3}$ \\ \hline
$10^{-7}$ &  $3.0\times 10^{-6}$& $12,451$ & $1.1\times 10^{-2}$ \\ \hline
$10^{-8}$ &  $4.2\times 10^{-7}$& $18,442$ & $1.7\times 10^{-2}$ \\ \hline
\end{tabular}
\caption{The relative residual norms of the approximate eigenpairs obtained
from $H_1$ matrices associated with different selection thresholds $\tau$,
the corresponding dimension of $H_1$, and the wall clock time required to
compute these approximations for the Heisenberg spin-1/2 Hamiltonian with $L=20$ spins.}
\label{tab:MBLconv}
\end{table}

When the greedy algorithm terminates, the dimension
of $H_1$ becomes 18,442, which is less than 10\% of the dimension of
$H$, which is 184,756 for $L = 20$ spins.
To illustrate the overall efficiency of this algorithm, we use the
\texttt{eigs} function
to compute the lowest eigenvalue and the corresponding eigenvector of
the full $H$ directly,
using a random vector as the starting guess.
The total wall clock time used in this full calculation
is more than seven times of that used by the greedy algorithm.
\begin{figure}[hbtp]
\centerline{%
\begin{subfigure}[t]{0.49\textwidth}
 \figname{mbl-conv-size}%
 \begin{tikzpicture}
 \begin{axis}[
   width=\textwidth,
   xmin=0, xmax=30,
   ymin=0, ymax=25000,
   scaled y ticks=false,
   xlabel={greedy iteration},
   title={matrix dimension $H_1$},
   legend pos=north west,%
   legend style={draw=none,fill=none},%
 ]
 \addplot[GreedyResStyle] table {\datfile{mbl-greedy-res-size}};
 \addplot[GreedyPertStyle] table {\datfile{mbl-greedy-pert-size}};
 \legend{\greedyres,\greedypert};
 \end{axis}
 \end{tikzpicture}%
\caption{The change in matrix dimension of $H_1$.}
\end{subfigure}%
\hfill%
\begin{subfigure}[t]{0.49\textwidth}
 \figname{mbl-conv-res}%
 \begin{tikzpicture}
 \begin{semilogyaxis}[
   width=\textwidth,
   xmin=0, xmax=30,
   ymin=1e-7, ymax=1e0,
   xlabel={greedy iteration},
   title={relative residual norm},
 ]
 \addplot[black,densely dashed] coordinates{(0,1e-6) (30,1e-6)};
 \addplot[GreedyResStyle] table {\datfile{mbl-greedy-res}};
 \addplot[GreedyPertStyle] table {\datfile{mbl-greedy-pert}};
 \end{semilogyaxis}
 \end{tikzpicture}%
\caption{The change in relative residual norm.}
\end{subfigure}%
}
\caption{A comparison of two versions of the greedy algorithm
that use different metric to select rows and columns of the disordered quantum spin chain Hamiltonian.}\label{fig:MBLcomp}
\end{figure}
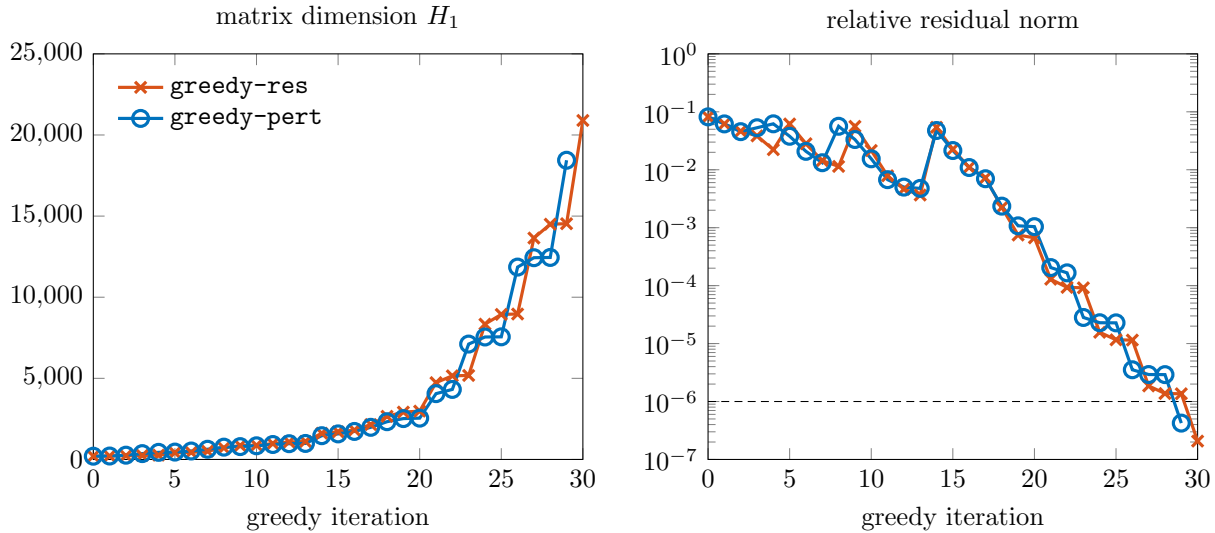

We then compare the perturbation analysis based greedy approach (\greedypert) to
the residual based (\greedyres) approach.  Although it may be seen, in
Figure~\ref{fig:MBLcomp}, that the relative residual norm of the approximate
eigenpairs and the dimension of the submatrix $H_1$ evolve similarly with the
number of greedy iterations in the two approaches, the small differences in the
dimension of $H_1$ yield slightly different time to solution.

We use the same dynamical threshold adjusting scheme to for select additional
rows and columns of $H$ in the first order componentwise perturbation method.

\subsection{California Road Network}
\label{sec:examples:network}

\begin{figure}[hbtp]
 \centering
 \hfill
   \begin{subfigure}[t]{.5\textwidth}
     \centering
     \includegraphics[width=0.9\textwidth]{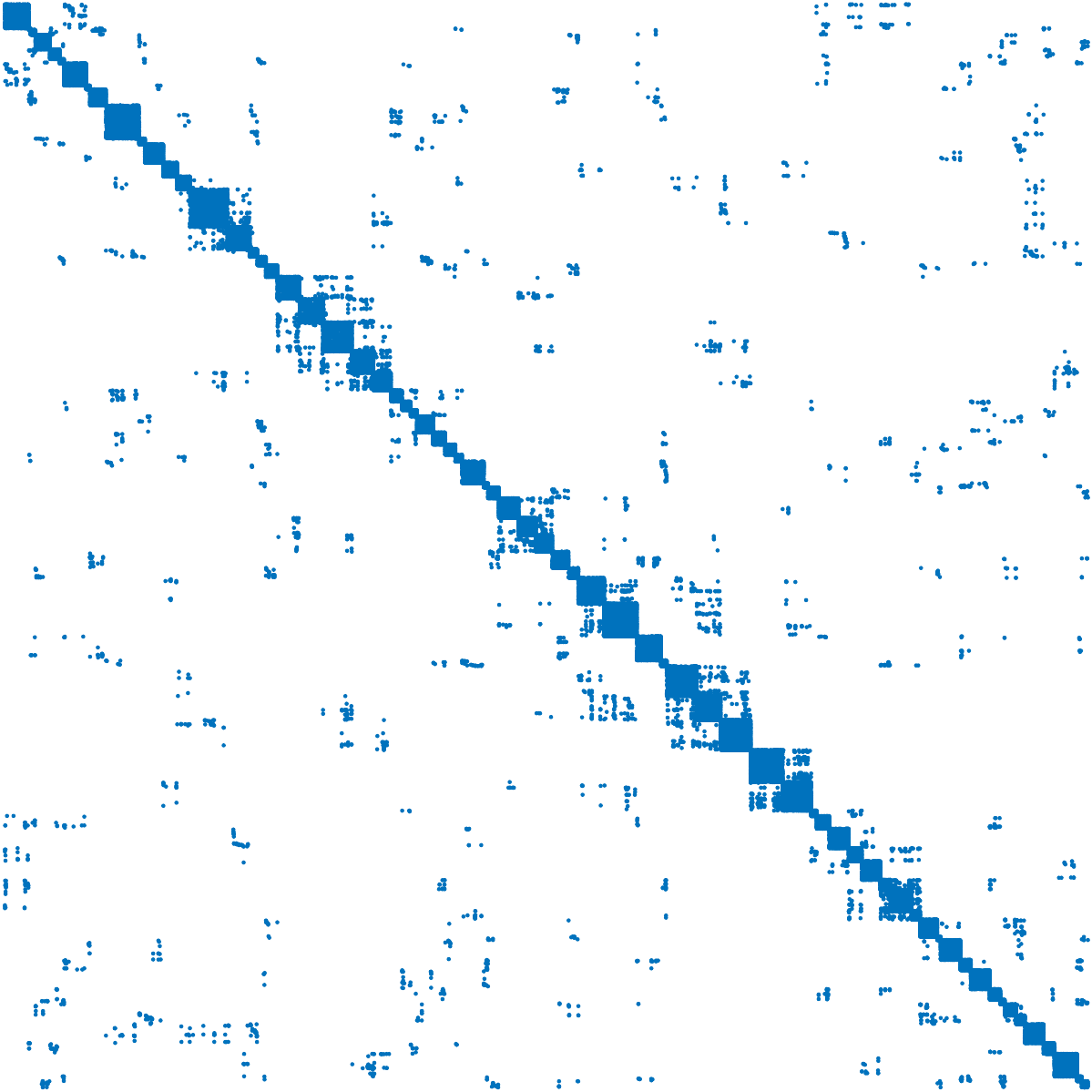}
     \caption{The sparsity pattern of the adjacency matrix.}
   \end{subfigure}%
 \hfill
    \begin{subfigure}[t]{.5\textwidth}
    \centering
    \figname{crn-eigvec1}%
    \begin{tikzpicture}[spy using outlines =
                        {magnification=15,size=0.4\textwidth,connect spies}]
    \begin{axis}[
      width=0.95\textwidth,
      height=0.95\textwidth,
      xmin=0, xmax=1971281,
      scaled x ticks=false,
      ymin=0, ymax=0.6,
      xtick={0,1000000,1971281},
      xticklabels={0,{1,000,000},{1,971,281}},
      extra x ticks={500000,1500000},
      extra x tick labels={},
      xlabel={$i$},
      ylabel={$|x(i)|$},
    ]
    \addplot+[ycomb,mark=none,myblue,line width=1pt]
      table {\datfile{crn-eigvec1}};
    \addplot[line width=0.1pt] coordinates {(498391,0) (498391,0.002)};
    \addplot[line width=0.1pt] coordinates {(534752,0) (534752,0.002)};
    \draw node[fill=white] at ( 971000,0.16) {\tiny 498,391};
    \draw node[fill=white] at (1517000,0.16) {\tiny 534,752};
    \coordinate (spypoint) at (axis cs:517000,0.0075);
    \coordinate (magnifyglass) at (axis cs:1250000,0.3);
    \end{axis}
    \spy[spy connection path={
      \begin{scope}[on background layer]
      \draw[very thin,densely dotted] (tikzspyonnode.north east) --
                                      (tikzspyinnode.north east);
      \draw[very thin,densely dotted] (tikzspyonnode.north west) --
                                      (tikzspyinnode.north west);
      \draw[very thin,densely dotted] (tikzspyonnode.south west) --
                                      (tikzspyinnode.south west);
      \draw[very thin,densely dotted] (tikzspyonnode.south east) --
                                      (tikzspyinnode.south east);
    \end{scope}
    }] on (spypoint) in node at (magnifyglass);
    \end{tikzpicture}
    \caption{Magnitude of each element of principal eigenvector.}
    \end{subfigure}
 \caption{The sparsity structures of the adjacency matrix associated with
          a California road network, and its principal eigenvector.}
 \label{fig:CAroad}
\end{figure}
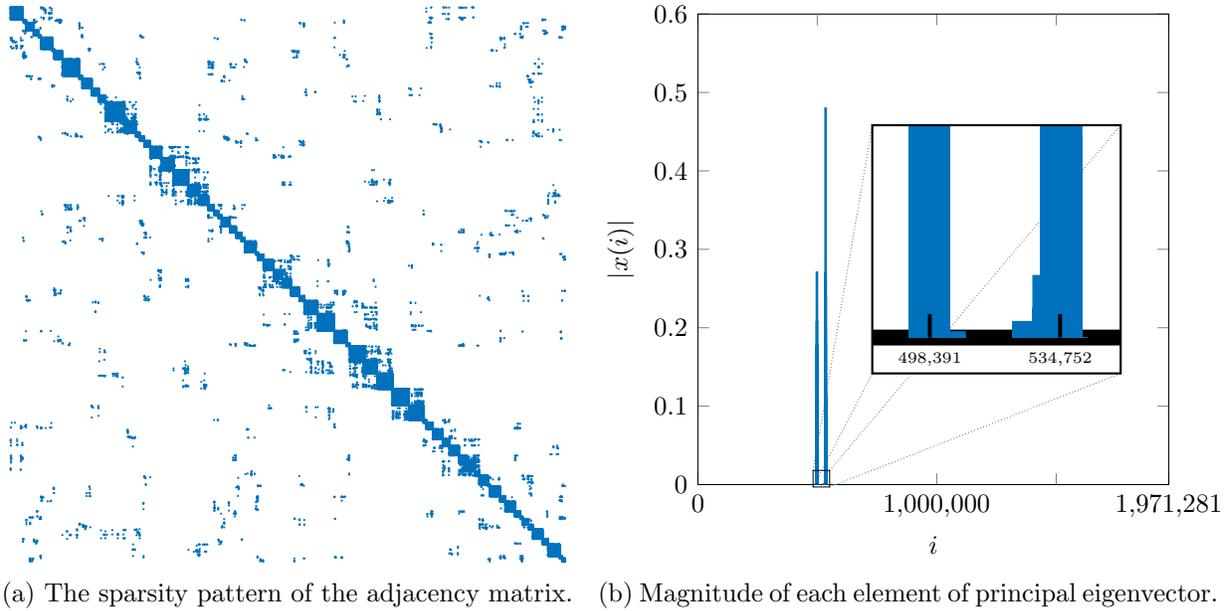

In this section, we demonstrate the efficiency of the greedy algorithm when it is
used to compute the principal eigenvector (PEV), i.e., the eigenvector associated
with the largest (in magnitude) eigenvalue, of an adjacency matrix $H$ representing
a California road network. This matrix is part of the Stanford Network Analysis Project
(SNAP) collection~\cite{snap} and can be download from the SuiteSparse Matrix
collection~\cite{suitesparse}. The dimension of $H$ is $n = 1,971,281$ and its sparsity
pattern is shown in Figure~\ref{fig:CAroad}(a). Each column (or row) of the matrix
corresponds to a road intersection or endpoint which is a vertex of the corresponding
adjacency graph. Each nonzero element of the matrix represents a connection between two
intersections or endpoints, which constitutes an edge between two vertices on
the adjacency graph.  The principal eigenvector of $H$ is localized as shown in Figure~\ref{fig:CAroad}(b).  Note that each of the two peaks in Figure~\ref{fig:CAroad}(b) actually correspond to
several nonzero components of the PEV. Due to the large dimension of this matrix,
the figure does not show how the nonzero elements are distributed around these
two peaks. The localization of the eigenvector often indicates a densely connected
subgraph. There has been a lot of effort in understanding the localization
of the PEV in terms of the degrees and centralities of the nodes and structure of the
graph theoretically~\cite{networkPEV,lapPEV,centrality}.
The greedy algorithm provides an efficient numerical validation tool for these theoretical
studies.

To compute the PEV of the adjacency matrix of the California road
network, we apply the same componentwise perturbation method
with an adaptively modified selection threshold that we used to compute the
ground state of the MBL Hamiltonian in section~\ref{sec:examples:mbl}
to incrementally select more rows and columns of $H$ to add to $H_1$.
However, instead of
computing the algebraically smallest eigenvalue of each submatrix identified
in the greedy algorithm, here we compute the eigenvalue with the largest
magnitude.  Furthermore, while in the MBL problem it suffices to select a
random subset of rows and columns of $H$ to construct the initial
approximation to the desired eigenvalue and eigenvector, here a more careful
selection of the initial approximation is needed.

Because it is believed that the nonzero components of the PEV corresponds
to a densely connected subgraph of the road network graph, we first
pick a row/column that has a large number of nonzeros.
This row/column corresponds to the node in the network with a large
degree. We then select rows and columns that correspond to nodes
connected to the initially selected node with a large degree.  All these
selected rows and columns form our initial $H_1$.
For the California road network,
there is one node that has the highest degree of 12. There are two nodes
with degree 10. All other nodes have degree less than 10.

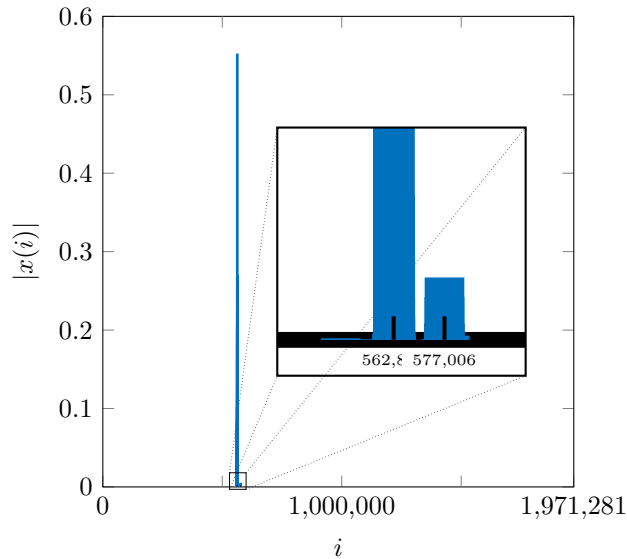
\begin{figure}[b!]
 \centering
    \figname{crn-eigvec2}%
    \begin{tikzpicture}[spy using outlines =
                        {magnification=15,size=0.2\textwidth,connect spies}]
    \begin{axis}[
      width=0.475\textwidth,
      height=0.475\textwidth,
      xmin=0, xmax=1971281,
      scaled x ticks=false,
      ymin=0, ymax=0.6,
      xtick={0,1000000,1971281},
      xticklabels={0,{1,000,000},{1,971,281}},
      extra x ticks={500000,1500000},
      extra x tick labels={},
      xlabel={$i$},
      ylabel={$|x(i)|$},
    ]
    \addplot+[ycomb,mark=none,myblue,line width=1pt]
      table {\datfile{crn-eigvec2}};
    \addplot[line width=0.1pt] coordinates {(562819,0) (562819,0.002)};
    \addplot[line width=0.1pt] coordinates {(577006,0) (577006,0.002)};
    \draw node[fill=white] at (1217000,0.16) {\tiny 562,819};
    \draw node[fill=white] at (1430000,0.16) {\tiny 577,006};
    \coordinate (spypoint) at (axis cs:565000,0.0075);
    \coordinate (magnifyglass) at (axis cs:1250000,0.3);
    \end{axis}
    \spy[spy connection path={
      \begin{scope}[on background layer]
      \draw[very thin,densely dotted] (tikzspyonnode.north east) --
                                      (tikzspyinnode.north east);
      \draw[very thin,densely dotted] (tikzspyonnode.north west) --
                                      (tikzspyinnode.north west);
      \draw[very thin,densely dotted] (tikzspyonnode.south west) --
                                      (tikzspyinnode.south west);
      \draw[very thin,densely dotted] (tikzspyonnode.south east) --
                                      (tikzspyinnode.south east);
    \end{scope}
    }] on (spypoint) in node at (magnifyglass);
    \end{tikzpicture}
 \caption{The magnitude of the eigenvector associated with the second largest
          (in magnitude) eigenvalue of the California road network adjacency matrix.}
 \label{fig:RoadCAev2}
\end{figure}

If we construct the initial $H_1$ by selecting rows and columns
corresponding to the node with the maximum degree as well as
all nodes connected to the maximum-degree node, which yields
a $13\times 13$ matrix, the greedy algorithm converges to the
second largest eigenvalue of $H$. Even though this is not
the largest eigenvalue, the corresponding eigenvector, which is shown in
Figure~\ref{fig:RoadCAev2}, is localized also with 798 large
elements in magnitude.  The localization region
is different from the PEV shown in Figure~\ref{fig:CAroad}(b).
The subgraph containing these nodes is potentially interesting to
examine also.

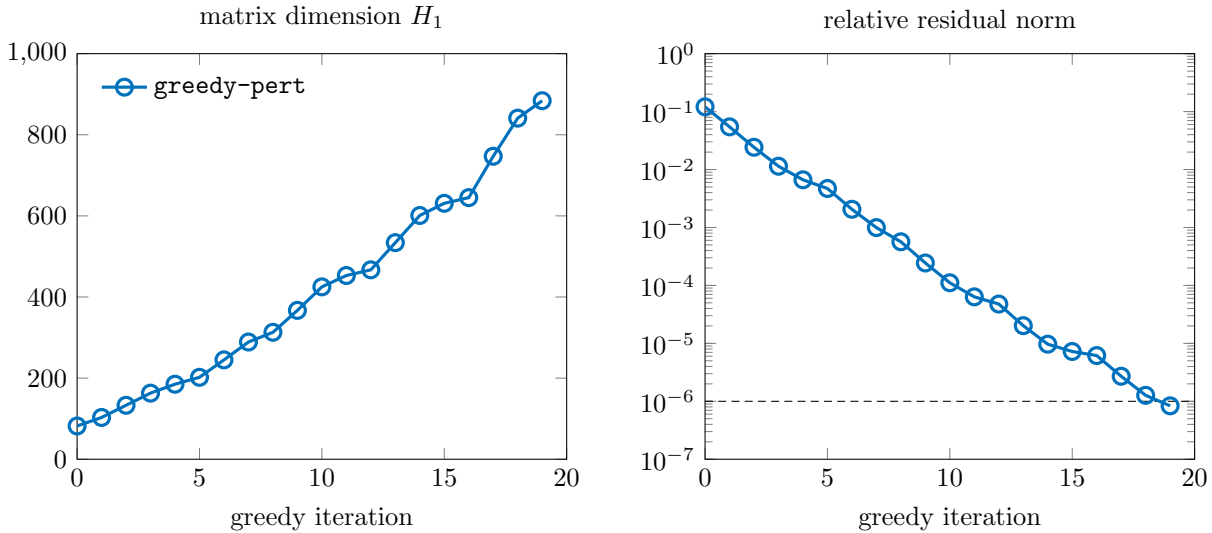
\begin{figure}[hbtp]
\centerline{%
\begin{subfigure}[t]{0.49\textwidth}
 \figname{crn-conv-size}%
 \begin{tikzpicture}
 \begin{axis}[
   width=\textwidth,
   xmin=0, xmax=20,
   ymin=0, ymax=1000,
   scaled y ticks=false,
   xlabel={greedy iteration},
   title={matrix dimension $H_1$},
   legend pos=north west,%
   legend style={draw=none,fill=none},%
 ]
 \addplot[GreedyPertStyle] table {\datfile{crn-greedy-pert-size}};
 \legend{\greedypert};
 \end{axis}
 \end{tikzpicture}%
\caption{The change in matrix dimension of $H_1$.}
\end{subfigure}%
\hfill%
\begin{subfigure}[t]{0.49\textwidth}
 \figname{crn-conv-res}%
 \begin{tikzpicture}
 \begin{semilogyaxis}[
   width=\textwidth,
   xmin=0, xmax=20,
   ymin=1e-7, ymax=1e0,
   xlabel={greedy iteration},
   title={relative residual norm},
 ]
 \addplot[black,densely dashed] coordinates{(0,1e-6) (30,1e-6)};
 \addplot[GreedyPertStyle] table {\datfile{crn-greedy-pert}};
 \end{semilogyaxis}
 \end{tikzpicture}%
\caption{The change in relative residual norm.}
\end{subfigure}%
}
 \caption{The convergence of the greedy algorithm applied to the California Road Network problem.}
 \label{fig:CAroadResults}
\end{figure}

In order to obtain the largest eigenvalue, we select the rows/columns
of the 3 nodes with degree 10 or higher, as well as all the corresponding
rows/columns associated with the nodes connecting to these 3 large degree
nodes up to distance 2.
The dimension of this initial $H_1$ is $82 \times 82$.
Figure~\ref{fig:CAroadResults} shows how the dimension of the submatrix $H_1$
and the relative residual norm of the approximate
eigenpair evolves with respect to the number of greedy iterations.

When the greedy algorithm terminates, the dimension of $H_1$ is 884,
which is roughly 0.04\% of the dimension of $H$.
When compared to computing the PEV of $H$ directly using the {\tt eigs}
function, the greedy algorithm is 420 times faster in terms of wall clock time.

\section{Conclusions}
\label{sec:conclude}

We presented a greedy algorithm for computing an
eigenpair of a symmetric matrix $H$ that has localization
properties. The key feature of the algorithm is to identify and select
rows and columns of $H$ to be included in a submatrix $H_1$ that
can be easily diagonalized (partially). The eigenvectors thus obtained for this submatrix are then
used to generate more efficient starting guesses for iterative diagonalization of the full matrix $H$ if the eigenvector to be computed is not strictly localized.
We discussed a
number of greedy strategies and criteria for such a selection, and
presented numerical examples using Hamiltonian matrices arising in two types of quantum many-body
eigenvalue problems as well as an adjacency matrix describing a California road network.
For all these problems, we found that the residual
based selection approach is almost as good as the strategy based on
perturbation analysis.  Both approaches require computing $Bx_1$,
where the submatrix $B$ is defined in \eqref{eq:Hpart}. For large problems,
accessing the entirety of $B$ may be prohibitively expensive, especially
when the nonzero matrix elements of $H$ are not explicitly stored, even
though this submatrix is typically sparse. Algorithms based on
choosing and evaluating selected rows of $B$ will need to be developed.
Similarly, it may also be prohibitively expensive to access the $C$ matrix
in a full perturbation analysis. Methods for identifying good approximations
to $C$ that are easy to compute also need to be developed.
The efficient implementation of these algorithms is problem dependent
and beyond the scope of this paper.

Even though we only showed how to compute either the smallest (algebraically)
or the largest (in magnitude) eigenpair of $H$, the greedy algorithm developed
here can be used to compute more than one eigenpair. If more than one eigenpair
needs to be computed, the initial choice of $H_1$ and the subsequent selection
scheme will need to take into account the presence of several eigenvectors with
different localization patterns. We will investigate the practical aspects of
the greedy algorithm for solving this type of problem in future work.

We should also note that there are other specialized techniques for computing
localized eigenvectors and the corresponding eigenvalues in specific
applications. Examples include the method based on finding a so-called landscape
function and its local maximizers proposed in~\cite{specpred} and
combinatorial approaches that can be used
to identify a dense clusters or subgraph of a network~\cite{subgraph}.
The greedy algorithm proposed in this paper is not meant to replace these
methods. However, it can be used to validate eigenvalues and eigenvector
obtained in these methods, and is therefore a useful and complementary tool.

\section*{Acknowledgments}
This work was supported in part by the U.S.~Department of
Energy, Office of Science, Office of Workforce Development for Teachers
and Scientists (WDTS) under the Science Undergraduate Laboratory
Internship (SULI) program, the U.S. Department of Energy,
Office of Science, Office of Advanced Scientific Computing Research, Scientific
Discovery through Advanced Computing (SciDAC) program, and
the U.S.~Department of Energy, Office of Science, Office of Nuclear Physics,
under Award Number DE-FG02-95ER-40934.

\bibliographystyle{abbrvurl}
\bibliography{references}

\end{document}